\documentclass{amsart}

\usepackage[utf8]{inputenc}
\usepackage[fleqn]{mathtools}

\usepackage{geometry}

\usepackage{amssymb,amsthm,mathrsfs}
\usepackage{dsfont}
\usepackage{graphicx}
\usepackage{pdfpages}
\usepackage{subcaption}
\usepackage{makecell}
\usepackage[ruled,vlined,linesnumbered,displayblockmarkers]{algorithm2e}
\usepackage{tikz}
\usepackage{bbm}
\usetikzlibrary{positioning}
\usepackage{tikz-cd}
\usepackage{float}
\usepackage{array}

\usepackage[british]{babel}
\usepackage[babel]{csquotes}

\usepackage[backend=biber, date=long, giveninits=true, terseinits=true, maxnames=10]{biblatex}
\bibliography{report.bib}
\usepackage{nameref}
\usepackage{enumitem}
\DeclareNameAlias{sortname}{family-given}
\DeclareNameAlias{author}{family-given}

\newtheorem{lemma}{Lemma}
\newtheorem{theorem}[lemma]{Theorem}

\theoremstyle{definition}
\newtheorem{definition}[lemma]{Definition}
\theoremstyle{remark}

\newtheorem*{remark*}{Remark}

\usepackage{fancybox}
\usepackage{todonotes}
\usepackage{menukeys}
\newcounter{todocounter}

\presetkeys{todonotes}{inline, color=blue!30}{}

\DeclareMathOperator{\RR}{\mathbb R}
\DeclareMathOperator{\NN}{\mathbb N}
\DeclareMathOperator{\ZZ}{\mathbb Z}

\DeclareMathOperator{\KK}{\mathcal K}

\usepackage[pdftex,
            pdfproducer={LaTeX},
            hidelinks,
            pdfcreator={pdflatex}]{hyperref}

\title{Detecting Temporal shape changes with the Euler Characteristic Transform}
\author[Marsh, Zhou, Qin, Lu, Byrne \& Harrington]{Lewis Marsh$^{1,2}$, Felix Y Zhou$^{2,3}$, Xiao Qin$^{2,4}$, Xin Lu$^2$, Helen M Byrne$^{1,2}$, Heather A Harrington$^{1,5}$}
\date{June 2022}
\address{$^1$Mathematical Institute, University of Oxford, Oxford, UK\\
$^2$Ludwig Institute for Cancer Research, University of Oxford, Oxford, UK\\
$^3$Lyda Hill Department of Bioinformatics, UT Southwestern Medical Center, Dallas, Texas, USA\\
$^4$Department of Oncology, University College London, London, UK
$^5$Wellcome Centre for Human Genetics, University of Oxford
}
\email{harrington@maths.ox.ac.uk}

\begin{document}

\maketitle

\begin{abstract}
    Organoids are multi-cellular structures which are cultured \emph{in vitro} from stem cells to resemble specific organs (e.g., brain, liver) in their  three-dimensional composition.  
    Dynamic changes in the shape and composition of these model systems can be used to understand the effect of mutations and treatments in health and disease.  
    In this paper, we propose a new technique in the field of topological data analysis for DEtecting Temporal shape changes with the Euler Characteristic Transform (DETECT). DETECT is a rotationally invariant signature of dynamically changing shapes. We demonstrate our method on a data set of segmented videos of mouse small intestine organoid experiments and show that it outperforms classical shape descriptors. We verify our method on a synthetic organoid data set and illustrate how it generalises to 3D. We conclude that DETECT offers rigorous quantification of organoids and opens up
    computationally scalable methods for distinguishing different growth regimes and assessing treatment effects.
\end{abstract}

\section{Introduction}

Organoids are three-dimensional multi-cellular \emph{in vitro} cell cultures that can be generated from stem cells. Organoids are also known as \emph{mini organs} because their constituent cells can differentiate into various cell lineages with defined cellular functions. 
At early times, their growth is approximately radially symmetric. Growth factors in the culture medium surrounding the organoids, drive the stem cells to proliferate and the organoids to increase in size. On longer times, the progeny of the stem cells differentiate or specialise, into different cell types that self-organise to produce tissues whose composition resembles the original organ. Excessive cell proliferation combined with cell differentiation can lead to complex shape changes, with the outer boundary of the organoids adopting intricate and asymmetric structures. 
The increasing use of organoids for studying 
tissue development and disease progression 
tissue responses to genetic and environmental perturbations can be attributed to their ability to recapitulate the 3D cellular architecture and function of their tissue of origin than 2D cell cultures \cite{kretzschmar2016organoids}. 
Organoids are also being used to investigate tissue responses to genetic and environmental perturbations and for drug testing and development \cite{Byrne2020}.
While organoids can now be generated to study multiple organs, including the brain, kidney and liver, the most widely studied organoids are those that derive from the intestinal epithelium, one of the fastest-renewing mammalian tissues.

Differentiated cells with specific cellular functions often exhibit defined cellular morphology.
Descriptors of morphology can be used as a phenotypic read-out of the organoids, which may reflect the underlying genetic composition and its response to environmental perturbations. Therefore, several studies have focused on quantifying and analysing organoid morphology. Simple measures, such as cell numbers, organoid volume and surface area,  diameter, shape factor (ratio of surface area to volume), and growth rate have been used to relate morphology, genotype and drug responses \cite{buske2012biomechanics, hartung2014mathematical, kim2020comparison,yan2018three}. Furthermore, deep learning methods can segment organoid images and extract morphological features including organoid perimeter and eccentricity \cite{gritti2021morgana, kassis2019orgaquant}.
At the same time, mechanistic models have been developed to investigate the relationship between stem cell proliferation, cell fate specification, organoid growth and morphology. These agent-based and continuum-based models have been compared with experimental data using
simple growth curves \cite{hartung2014mathematical,Thalheim2018,yan2018three}.
The complexity of organoid morphology lends itself to more sophisticated analysis. For example, genus and average curvature \cite{ishihara2022topological}, measures from geometry and topology, have been used to distinguish shape changes in organoids. Here we propose studying the geometry and topology of organoids with topological data analysis.

Topological data analysis (TDA) is a collection of data science methods for quantifying shape in data sets by using techniques from topology. TDA has been successfully applied to a variety of problems in applied mathematics, which include sensor network coverage \cite{de2007coverage}, image signalling \cite{boche2013signal, chung2009persistence}, skeletonising of images  \cite{delgado2014skeletonization}, discovering periodicities in time-series \cite{perea2015sw1pers}, material science \cite{kramar2013persistence, kramar2014quantifying}, and financial mathematics \cite{gidea2017topology,leibon2008topological}. A field of applied mathematics in which TDA has been particularly successful is mathematical biology, for example in studying enzyme kinetics \cite{marsh2022algebra}, neuron morphology \cite{kanari2018topological,beers2022stability} and protein structures \cite{dey2018protein,gameiro2015topological, martino2018supervised}.

One technique in TDA, introduced by Turner, Mukherjee and Boyer \cite{Turner2014}, is the Euler Characteristic Transform (ECT). The ECT gives a signature for shapes embedded in Euclidean space~$\RR^d$. 
The ECT and its extensions have been applied to a range of problems including   MRT scans of glioblastoma patients \cite{Crawford2020,jiang2020weighted}, to classification of 3D shapes (e.g. mammalian teeth \cite{wang2021statistical}, protein structures \cite{tang2022topological} and barley seeds \cite{Amezquita2021}). The above studies highlight the insight that ECT can generate when applied to static data. Here we show the additional information that can be gained from generalising the ECT to dynamic shape changes (i.e., spatio-temporal data).

We propose an extension to the smooth ECT (SECT), \emph{DEtecting Temporal shape changes with the Euler Characteristic Transform} (DETECT) which is the first application of a rotationally invariant and dynamic ECT. 
The output from DETECT is a signature in a Hilbert space that can be thought of as a surface in $\RR^3$.
To aid classification of the resulting signature, we apply a non-linear transformation using kernel methods. 
Other studies have successfully used kernels in conjunction with the ECT \cite{Crawford2020, tang2022topological, wang2021statistical}; we are the first to use approximate feature embeddings with the Nystroem method \cite{williams2000using}. In practice, Nystroem approximations typically scale better in runtime and memory than kernel methods \cite{williams2000using}, including Gaussian process approaches with the ECT \cite{Crawford2020, tang2022topological, wang2021statistical}. Compared with standard kernel methods, these approximations also allow a wider variety of methods to be used with DETECT, including random forests. 
We apply DETECT to dynamic data describing the boundaries of experimental and synthetic organoids in 2D and 3D.
We demonstrate the proposed method on the applicability of both synthetic and experimental on 2D and 3D data to distinguish morphology of organoids. First, we show that the smooth Euler Characteristic Transform super-cedes standard measures used to study organoid morphology. Next, we show that the temporal method proposed, DETECT, improves the classifying of different organoid treatment groups. 
Our paper is organised as follows. We first introduce relevant background on topological data analysis and kernel methods. We then introduce a novel signature, DETECT, which quantifies the evolution of a shape over time. By applying DETECT to a data set of 2D organoid boundaries segmented from videos of mouse small intestine organoid experiments, we demonstrate that DETECT can distinguish between treated and untreated organoids. Our results thereby give insights into how cancer treatments affect the morphology of organoids. We highlight that DETECT outperforms classical shape descriptors at this classification task. Finally, we demonstrate on a data set of 3D organoid boundaries generated by a mechanistic model of Yan et al. \cite{yan2018three} that our method generalises to 3D boundaries and verify that it can extract biologically meaningful information in a setting in which we have precise control over all biologically meaningful factors.

\section{Mathematical methods}

\subsection{Preliminaries}

\subsubsection{Simplicial complexes and filtrations}

To undertake our work, we require a mathematical definition of a shape. In topological data analysis (TDA) and computational geometry, simplicial complexes are widely used for this purpose:

\begin{definition}
Given a finite set of vertices $X$, an \emph{abstract simplicial complex} is a set of subsets of $X$, denoted $\KK$, such that for any $\tau\in\KK$ and $\sigma\subseteq\tau$, we have $\sigma\in\KK$. We call $\sigma\in\KK$ a \emph{simplex} of $\KK$. Moreover, for any simplex $\sigma$, we define $\dim(\sigma)=|\sigma| - 1$ and $\KK_i=\{\sigma\in\KK\,\vert\,\dim(\sigma)=i\}$.

A \emph{geometric (or embedded) simplicial complex} $\KK$ is an abstract simplicial complex that is endowed with an embedding in $\RR^d$. That is, $X\subset\RR^d$ and for all $\sigma,\tau\in\KK$ with $\sigma\neq\tau$ we have $\mathrm{relint}(\mathrm{cvx}(\sigma))\cap\mathrm{cvx}(\tau)=\varnothing$. Here, $\mathrm{relint}$ is the relative interior\footnote{For a convex set $C\subset\RR^d$, that is $\{x\in C\,\vert\,\forall y\in C:\exists \lambda>1:\lambda x + (1-\lambda)y\in C\}$.} and $\mathrm{cvx}$ the convex hull.\footnote{That is, the intersection of all convex subsets of $\RR^d$ that contain the given set of points.} We can then think of $\KK$ equivalently as the union of the convex hulls of all of its simplices, which is a topological subspace of $\RR^d$.

Given two simplicial complexes $\KK$ and $\KK'$, a \emph{simplicial map} $f$ is a function $f:X\to X'$, extending to a map $f:\KK\to\KK'$ by $f(\sigma)=\left\{f(x)\,\vert\,x\in\sigma\right\}$.
\end{definition}

\begin{figure}
\centering
    \includegraphics[width=0.15\textwidth]{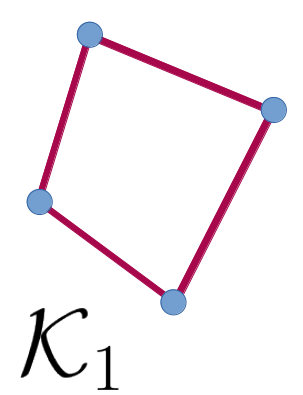}\hspace{2cm}%
    \includegraphics[width=0.15\textwidth]{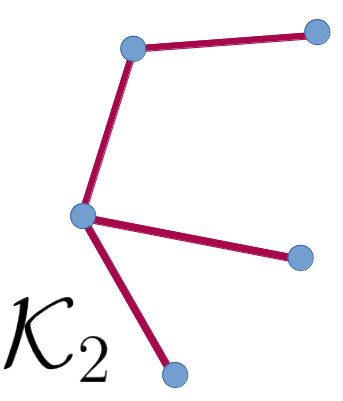}
\caption[Example of simplicial complexes]{Example of two simplicial complexes, $\mathcal{K}_1$ and  $\mathcal{K}_2$, embedded in $\RR^2$. Vertices as blue dots, 1-simplices as red lines.}
\end{figure}

Note that any abstract simplicial complex can be viewed as a geometric simplicial complex by considering $V$ to be a subset of the free vector space generated by $V$. This geometric simplicial complex is called the \emph{geometric realisation} of $\KK$.

In most settings, abstract simplicial complexes are more amenable to computations while geometric simplicial complexes, perhaps unsurprisingly, contain geometric information. The organoid boundaries extensively studied in the following section are all equivalent when modelled as abstract simplicial complexes. We use filtrations to study and compare geometric simplicial complexes representing organoid boundaries.

\begin{definition}
A \emph{filtration} of a simplicial complex $\KK$ is a function $f:\KK\to\RR$ such that $f(\sigma)\leq f(\tau)$ for $\sigma\subseteq\tau$. For $j\in\RR$, we then define
$$\KK^j=\left\{\sigma\in\KK\,\vert\, f(\sigma)\leq j\right\}.$$
All $\KK^j$ are simplicial complexes in their own right and $\KK^j\subseteq\KK^i$ for $j\leq i$.
\end{definition}

\subsubsection{Kernels and kernel approximations}

Before introducing further topology, we will briefly discuss kernels and their associated Hilbert spaces - a theory we will use when defining key topological signatures. 
Kernels are generalisations of inner-products. The motivation for generalising inner-products is twofold.   First, we may want to use data science methods requiring an inner-product (e.g. support vector classification, principal component analysis or k-means) on data in spaces not endowed with an inner-product. Second, even if it is possible to define an inner-product in data space, features in the data may not be linear. For example, not every labelled data set can be separated by a plane, resulting in inaccurate classification. In such an instance, SVC in combination with kernels could, by contrast, allow a separation. 

\begin{definition}
Let $\mathcal{X}$ be a non-empty set. A function $k:\mathcal{X}\times\mathcal{X}\to\RR$ is called a \emph{kernel} if it is symmetric and positive definite. That is, for any $x_1,...,x_n\in\mathcal{X}$ and $a_1,...,a_n\in\RR$, we get
$$\sum_{i,j=1}^na_ia_jk\left(x_i,x_j\right)\geq 0.$$
\end{definition}

For general $\mathcal{X}$, the Kronecker delta gives a kernel. If $\mathcal{X}$ is an inner-product space, the inner-product is a kernel. If $\mathcal{X}$ is a metric space with metric $d$, then the function
$$k(x,y):=\exp\left(-\frac{d(x,y)^2}{\lambda}\right),$$
where $\lambda>0$ is a hyperparameter, is a kernel called the \emph{Gaussian kernel}, which we employ in this work.

\begin{definition}
Let $k$ be a kernel on some set $\mathcal{X}$. If we define $\mathcal{H}_0=\mathrm{span}_{\RR}\{k(x,\,\cdot\,)\,\vert\,x\in\mathcal{X}\},$ then $\mathcal{H}_0$ is a vector space of functions from $\mathcal{X}$ to $\RR$. Note that
$$\langle k(x,\,\cdot\,),k(y,\,\cdot\,)\rangle_{\mathcal{H}_0}:=k(x,y)$$
defines an inner-product on $\mathcal{H}_0$ by bi-linear extension. We define $\mathcal{H}=\overline{\mathcal{H}_0}$, the completion of $\mathcal{H}_0$. Then $\mathcal{H}$ is a Hilbert space, called the \emph{Reproducing Kernel Hilbert Space} (RKHS) of $k$.
\end{definition}

The name RKHS derives from the fact that for any $f\in\mathcal{H}$ and $x\in\mathcal{X}$, we have $\langle k(x,\,\cdot\,), f\rangle_\mathcal{H}=f(x)$, which is called the reproducing property of $k$. In particular, the reproducing property of an RKHS together with the Cauchy-Schwarz inequality gives that the linear functionals $\delta_x:\mathcal{H}\to\RR$ defined by $\delta_x(f)=f(x)$ at $f\in\mathcal{H}$ are continuous for all $x\in\mathcal{X}$.

The main point - in the context of this work - of defining an RKHS is to illustrate that applying a kernel to elements of $\mathcal{X}$ can be viewed as first embedding $\mathcal{X}$ into some Hilbert space of functions $\mathcal{H}$ (by $x\mapsto k(x,\,\cdot\,)$) and then taking an inner-product of such embedded elements. While $\mathcal{H}$ may be infinite-dimensional and thus the embedding of $\mathcal{X}$ into $\mathcal{H}$ is intractable in general, we never need to compute the (exact) embedding itself - computing $k(x,y)$ for all $x,y\in\mathcal{X}$ is sufficient for any downstream method relying on the inner-product of $\mathcal{H}$ only. This insight is called the \emph{kernel trick}.

If $\mathcal{X}$ is of finite size $n$, then computing $k(x,y)$ for all pairs $k(x,y)$ may require only finitely many computations and will scale as $\mathcal O(n^2)$. To enable computations for large $n$, a number of approximation methods of lower computational complexity have been developed. One such method is the \emph{Nystroem approximation}:

\begin{definition}
Let $\mathcal{X}=\{x_1,...,x_n\}$ be of finite size $n$ and let $m< n$, where $m$ is an integer. Denote by $K$ the $n\times n$-matrix with $i,j$-entry $k(x_i,x_j)$. Then $K$ can be written in block-form as
$$
K=
\begin{bmatrix}
K_{11} & K_{12} \\
K_{21} & K_{22}
\end{bmatrix}.
$$
For $C:=[K_{11}, K_{12}]^T$, the $m$-th Nystroem approximation of $K$ is the matrix
$$\Tilde{K}=CK_{11}^\dagger C^T,$$
where ${}^\dagger$ denotes the Moore-Penrose pseudo-inverse.
\end{definition}

If $K$ is of approximate rank $m$ (or less), then $\Tilde{K}\approx K$ \cite{kumar2012sampling}. Typically, the approximate rank of a Gram matrix is much less than $n$ if $n$ is large. Computing $\Tilde{K}$ only requires $\mathcal O(mn)$ evaluations of $k$, where $m$ is typically fixed. In practice, we only compute $C(K_{11}^\dagger)^{1/2}$ (to save computer memory), as the standard inner-product of the $i$-th and $j$-th rows of $C(K_{11}^\dagger)^{1/2}$ approximately gives $k(x_i, x_j)$. This matrix $C(K_{11}^\dagger)^{1/2}$ can be used as a non-linear transformation and can be further analysed. In particular, we take the row vectors and feed them to a method using inner-products. Computing $C(K_{11}^\dagger)^{1/2}$ has a runtime complexity of $\mathcal O(nm^2+m^3)$. There exist sampling heuristics for picking an optimal set of $m$ points from $\mathcal{X}$ \cite{kumar2012sampling}.

\subsubsection{The Euler Characteristic and the Euler Characteristic Transform}

The \emph{Euler characteristic} is a topological invariant of simplicial complexes. Any two simplicial complexes that are (homotopy) equivalent as topological spaces have the same Euler characteristic. Conversely, if two simplicial complexes have different Euler characteristics, we can conclude that they are topologically different (i.e. not homotopy equivalent). We can compute the Euler characteristic entirely from the combinatorial information of an abstract simplicial complex.

\begin{definition}
Let $\KK$ be a simplicial complex. Then its Euler characteristic is
$$\chi(\KK)=\sum_{i=0}^\infty(-1)^i\cdot|\KK_i|.$$
\end{definition}

Given a simplicial complex $\KK$ embedded in $\RR^d$, using a sequence of Euler characteristics induced by a filtration, yields additional discriminative information to $\chi(\KK)$: Let $v\in S^{d-1}=\left\{x\in\RR^d:\|x\|_2=1\right\}$ be a fixed direction in $\RR^d$. We then call the filtration on $\KK$ induced by
$$f_v:\KK\to\RR,\quad\sigma\mapsto\min_{x\in\sigma}\left\{\langle x,v\rangle\right\}$$
the sub-level set filtration of $\KK$ in direction $v$, where $\langle\,\cdot\,,\,\cdot\,\rangle$ is the standard inner-product in $\RR^d$. We denote the above filtration $\KK^{\langle\,\cdot\,,v\rangle}$ and each sub-level-set at $t\in\RR$ as $\KK^{\langle\,\cdot\,,v\rangle\leq t}$.

Assume that $a\in\RR$ is larger than the diameter of $\KK$ in $\RR^d$. Then $f_v^{-1}((-\infty, -a))=\varnothing$ and $f_v^{-1}((-\infty, a))=\KK$ for all $v\in S^{d-1}$. We can now define  the Smooth Euler Characteristic transform and related constructions \cite{Turner2014}:
\begin{definition}
First, let the \emph{Euler characteristic curve} in a fixed direction $v$ be 
$$\mathrm{ECC}_{\KK}^v:[-a,a]\to\ZZ,\qquad t\mapsto\chi\left(\KK^{\langle v,\,\cdot\,\rangle\leq t}\right).$$
Secondly, this curve is smoothed by defining the \emph{smooth Euler characteristic curve} (SEC) as follows:
$$\mathrm{SEC}_{\KK}^v:[-a,a]\to\RR,\qquad t\mapsto\int_{-a}^t \left(\mathrm{ECC}_{\KK}^v(x)-\overline{\mathrm{ECC}_{\KK}^v}\right)\,\mathrm{d}x,$$
where $\overline{\mathrm{ECC}_{\KK}^v}$ is the mean of the function $\mathrm{ECC}_{\KK}^v$ over the interval $[-a,a]$.

Thirdly, the ECT of a shape $\KK$ is then the map
$$\mathrm{ECT}(\KK):S^{d-1}\to CF([-a,a]),\qquad v\mapsto\mathrm{ECC}_{\KK}^v,$$
where $CF([-a,a]$ is the set of constructible integer-valued functions on $[-a,a]$. For a more comprehensive discussion of these constructible functions and ECT, see \cite[5]{Curry2019}.
Finally, we can define the \emph{smooth Euler characteristic transform} (SECT):
$$\mathrm{SECT}(\KK):S^{d-1}\to {L}^2([-a,a]),\qquad v\mapsto\mathrm{SEC}_{\KK}^v.$$
Here, $L^2([-a,a])$ is the Hilbert space of all functions on $[-a,a]$ that can be generated by Fourier series.
\end{definition}

Turner and colleagues have shown that both the ECT and SECT are injective, making them sufficient statistics for comparing shapes embedded in $\RR^2$ and $\RR^3$ \cite{Turner2014}. Ghrist et al. \cite{ghrist2018persistent} and Curry et al. \cite{Curry2019} independently extended this injectivity result to general $\RR^d$. The ECT or SECT therefore also discriminate between shapes that are equivalent up to translation, rotation, reflection, and combinations thereof. The issue of discriminating between shapes equivalent up to translation can be overcome by re-centring simplicial complexes by subtracting the mean of all vertices from each vertex in the simplicial complex. However, resolving rotation and reflection requires more care. 
Fortunately, Curry et al. \cite{Curry2019} present a variant of the ECT that is injective on the space of shapes modulo actions of the orthogonal group $O(d)$.

Before introducing their result, we recall the notion of a pushforward measure:
\begin{definition}
Let $(X_1,\Sigma_1)$ and $(X_2,\Sigma_2)$ be measurable spaces, $f:X_1\to X_2$ be a measurable function, and $\mu:\Sigma_1\to[0,\infty]$ be a measure on $X_1$. Then $f_*\mu$, the pushforward of $\mu$ along $f$, is the measure on $X_2$ defined by $(f_*\mu)(U)=\mu(f^{-1}(U))$ for each $U\in\Sigma_2$.
\end{definition}
Then Theorem 6.6 in \cite{Curry2019} states:

\begin{theorem}\label{thm:rinv}
Let $\KK$ and $\KK'$ be generic simplicial complexes embedded in $\RR^d$. Let $\mu$ be the Lebesgue measure on $S^{d-1}$. If $\mathrm{ECT}\left(\KK\right)_*(\mu)=\mathrm{ECT}\left(\KK'\right)_*(\mu)$, then there exists a $\phi\in O(d)$ such that $\KK = \phi\left(\KK'\right)$.\label{thm:curry}
\end{theorem}

Note that the converse implication of the above theorem is trivial, as the Lebesgue measure on $S^{d-1}$ is invariant under the action of $O(d)$. Note that the function mapping ECC to SEC is injective. Therefore, Theorem \ref{thm:curry} generalises to the SECT.

For any $x\in[-a,a]$, define $\delta_x:{L}^2([-a,a])\to\RR$ by $\delta_x(f)=f(x)$ for all $f\in {L}^2([-a,a])$. These functionals, called evaluation functionals, are continuous and, hence, measurable as $L^2([-a,a])$ is an RKHS. Then if two embedded simplicial complexes, $\KK$ and $\KK'$ say, satisfy $K=\phi(\KK')$ for some $\phi\in O(d)$, we get
$$\int_{S^{d-1}}\delta_x\circ\mathrm{SECT}(\KK)\,\mathrm{d}\mu=\int_{{L}^2([-a,a])}\delta_x\,\mathrm{d}(\mathrm{SECT}(\KK)_*(\mu))$$
$$=\int_{{L}^2([-a,a])}\delta_x\,\mathrm{d}(\mathrm{SECT}(\KK')_*(\mu))=\int_{S^{d-1}}\delta_x\circ\mathrm{SECT}(\KK')\,\mathrm{d}\mu$$
for all $x\in[-a,a]$ by the change of variable formula for integrals with measure pushforwards and Theorem \ref{thm:curry}. Hence, the mean of such SECT evaluations, or collections thereof, form a statistic that can be used for distinguishing shapes modulo $O(d)$ actions.

\begin{figure}
    \centering
    \includegraphics[width=\textwidth]{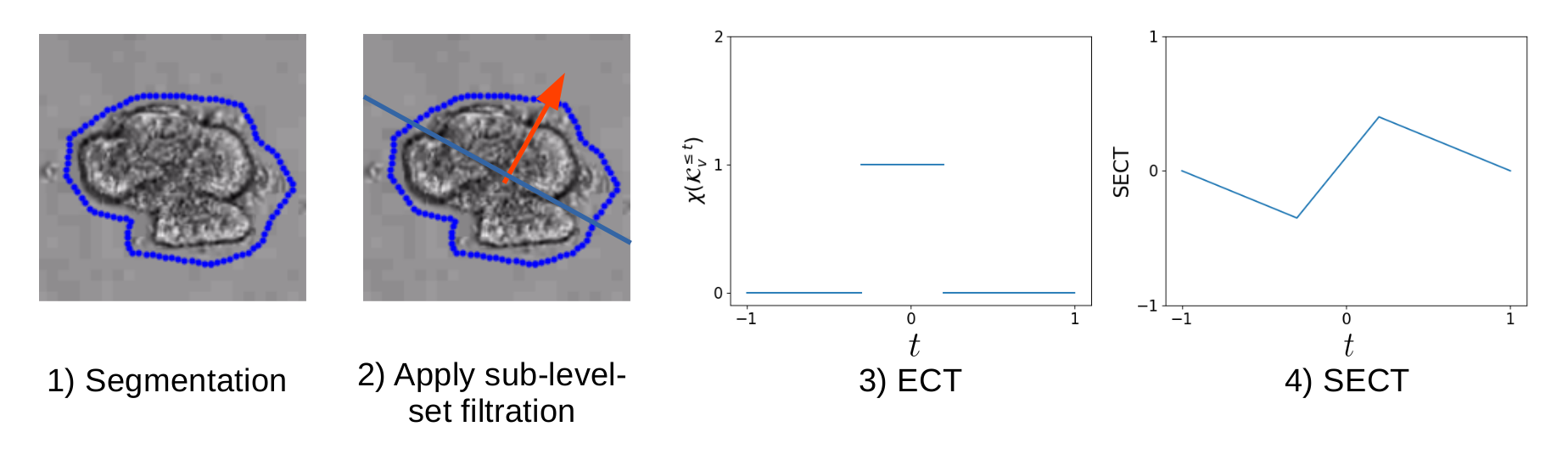}
    \caption{Standard ECT pipeline visualised on segmented organoid boundaries. 1) Segmented input data. 2) Illustration of the sub-level-set filtration in the direction given by the arrow. 3) The ECT in the direction given in 2).  4) The SECT in the direction given in 2).}
    \label{fig:ECT}
\end{figure}

\subsubsection{Previous studies using ECTs and kernel methods}

Kernel methods have been used successfully in conjunction with the SECT for both shape regression and classification problems \cite{Crawford2020, wang2021statistical, tang2022topological}. Both of these studies use Gaussian process models. Gaussian processes include the inversion of a Gram matrix and thus have a runtime of $\mathcal O(n^3)$. Hence, the Nystroem method we employ scales better to large data sets. These models are also conceptually more complex than linear regression and SVC, which we use for regression and classification. To the best of our knowledge, neither the ECT nor SECT have previously been used in combination with kernel approximation methods.

\subsection{Temporal shape detection}
The SECT transforms a fixed, static shape. We now extend the definition of the SECT to get a rotationally invariant temporal signature of a sequence of shapes:

\begin{definition}
Let $T=[0, c]$ or $T=\{0,1,...,n-1\}$, where $c\in\RR$ and $n\in\NN$, be a set of time points. Let $\{\KK (t)\}_{t\in T}$ be sequence of shapes. Then the  \emph{DETECT} (DEtecting Temporal shape changes with the Euler Characteristic Transform) of this sequence is the transform
 
$$\mathrm{DETECT}\left(\{\KK (t)\}\right):T\to L^2([-a,a]), \qquad t\mapsto\left(x\mapsto\int_{S^{d-1}}\delta_x\circ\mathrm{SEC}^v_{\KK (t)}\,\mathrm{d}v\right).$$
\end{definition}

Here, we will use DETECT with $T=\{0,...,n-1\}$.
In practice, after integrating out the dependence on the direction, DETECT is a continuous function from $T$ to $L^2([-a,a])$ (using the continuity of $\delta_x$ and that $L^2([-a,a])$ is an RKHS). Such functions form an infinite dimensional vector space and thus a finite presentation is not possible in general. We, therefore, evaluate $\mathrm{DETECT}$ at any fixed $t\in T$ on a finite number of evenly spaced points $P$ in $[-a,a]$. $\mathrm{DETECT}$ is then represented approximately by a $|T|\times |P|$-matrix. In this paper, we apply $\mathrm{DETECT}$ to two time-course data sets of organoid boundaries. We will consider the space of such matrices to be endowed with the $\|\,\cdot\,\|_2$-norm.

\section{Data sets}\label{sec:data}
We analyse experimental 2D video data as well as synthetic 3D spatio-temporal data. 

\subsection{Intestinal organoids from 2D video experiments}

We first acquired a set of imaging data derived from time-lapse imaging of mouse small intestine organoids.  
In total, we have 176 organoids and 320 video frames for each organoid. The data set comprises of 74 wild-type (WT) and 102 p53 knock-out (KO/mutant) genetics organoids. Both groups of organoids further split into untreated (CNT) and organoids treated with valproic acid and GSK3 inhibitor CHIR99021 (VC).
These organoids have been filmed throughout their growth and the resulting videos have been segmented. After segmentation, we have 100 points summarising the boundary of each organoid at each video frame (frames starting at beginning of the experiment and are taken every 15 min. henceforth).  We discard some videos which for technical reasons have fewer than 320 video frames (as most videos below this threshold still seem to change their morphology at the end of the video).  

\begin{figure}
    \centering
    \includegraphics[width=\textwidth]{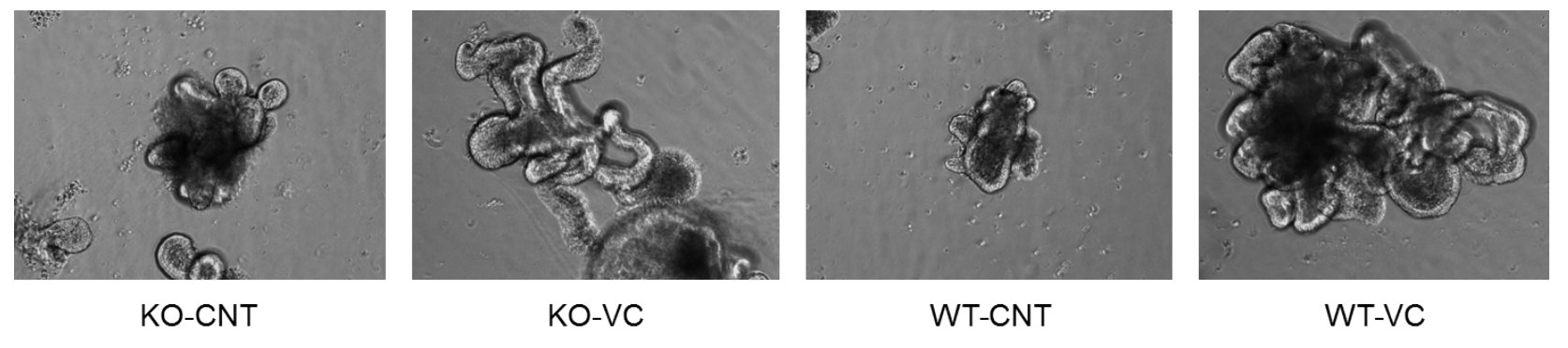}
    \caption{Phenotype effect of VC treatment to intestinal organoids. Static video snapshots of the final frame. One example is shown for each condition.}
    \label{fig:growth_eg}
\end{figure}

The organoid boundaries in this data set are, in the 2D video view, close to being perfectly circular in the early stages of the videos across all experimental conditions. This simple geometry is a result of the cellular homogeneity in the early phases.
As time progresses, cells proliferate and differentiate. Through proliferation, organoids grow in size, and through stem-cell differentiation, the cellular composition of organoids changes. As different cell types have different mechanistic properties and differentiation is not spatially uniform, organoids cease to be spherical. Most notably, they elongate and their boundary buckles, possibly leading to the growth of finger-like protrusions. Such growth behaviour is illustrated by the examples of final video frames presented in Figure \ref{fig:growth_eg}.

Each collection of boundary points is transformed into a simplicial complex representing the organoid boundaries, yielding a sequence of simplicial complexes indexed by $t=0,...,319$ for each organoid.
We re-centre each simplicial complex such that the mean of all vertices is the origin. We then compute the radius of the simplicial complex at $t=0$ (i.e. the largest norm of all vertices after re-centring) and divide all vertices in the sequence of simplicial complexes by that value. We translate and scale the data in this way to simplify it, given its limited size. As a result, the initial size of organoids or any movement throughout time is not considered by any downstream analysis, including DETECT.

\subsection{Spatio-temporal synthetic data}

We next apply DETECT to a synthetic data set to verify our findings on the experimental data. We use data generated by a model first presented in Yan et al. \cite{yan2018three}. Yan et al. construct a mechanistic model describing the growth of cancerous colon organoids containing stem, progenitor and terminally differentiated cells in 3D. These cells differ in terms of their rates of mitosis, differentiation and cell death, the rates at which these processes occur are controlled by model parameters. 

\begin{figure}
    \centering
    \includegraphics[width=\textwidth]{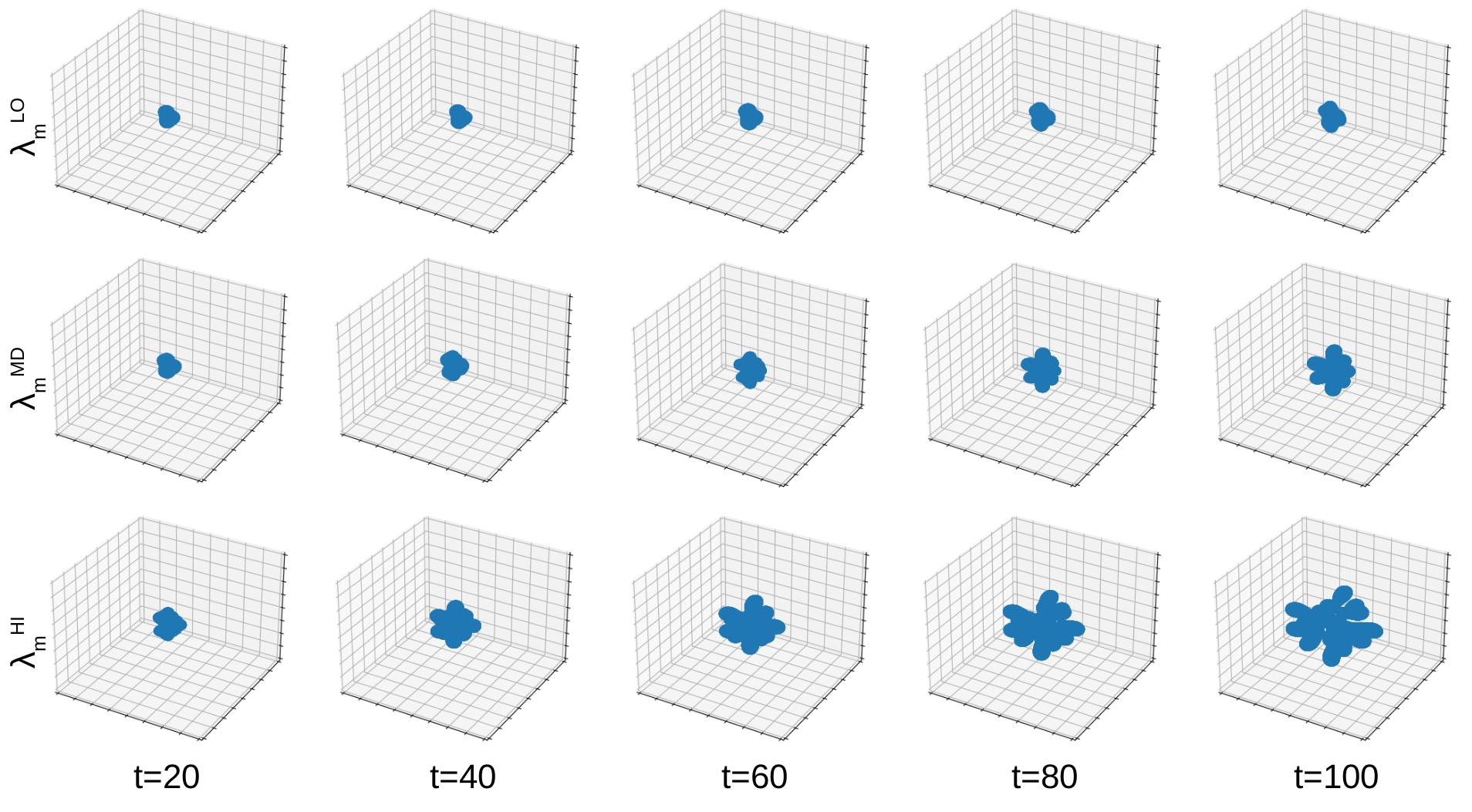}
    \caption{Spatio-temporal shape changes in 3D of synthetic organoids. Top column: Low (LO) mitosis rate. Middle column: Medium (MD) mitosis rate. Bottom column: High (HI) mitosis rate. In all organoids visualised, the lysis rates are at the lowest values given in the data set.}
    \label{fig:syn_time_course}
\end{figure}

We fix all but two of the parameters presented in \cite{yan2018three}. We vary $\lambda_L=0.5,\, 1,\, 2$, a death-rate parameter, and $\lambda^\mathrm{SC}_m=\lambda^\mathrm{CP}_m=0.35,\, 0.71,\, 1.42$, cell mitosis parameters for stem cells and committed progenitor cells, respectively (all parameters are dimensionless). All other parameters are left at the default values given in \cite{yan2018three}. As we observe all possible combinations of death and proliferation parameters, we get time-course data for nine different computationally modelled organoids. We visualise examples of the simulated organoid development in this data set in Figure \ref{fig:syn_time_course}.

Unlike the experimental data, the number of (3D) boundary points varies proportionally with the size of the simulated organoid. To ensure computational tractability, we restrict ourselves to 300 boundary points sampled uniformly at random at each time point. Different 300 point samples do not lead to any notable perturbations in the downstream analyses.
To triangulate the boundary surface, we first re-centre the organoid such that the mean of all boundary points is the origin. We then perform a stereographic projection into the xy-plane and perform a Delaunay triangulation and identify those points bordering an infinite area 2-cell. After projecting the finite components of the triangulations back onto the sphere, we add a further point, which is the mean of all points bordering an infinite area cell in the previous step, and insert 1 and 2-cells to fill the north-pole area of our organoid. We credit \cite{preproc} with this pre-processing procedure. Unlike in the experimental data, re-scaling is not needed as all simulated organoids are identical at $t=1$.

We perform this pre-processing step to ensure that the resulting simplicial complex has a geometric realisation homeomorphic to a sphere (or a union thereof, if an organoid disconnects). This pre-processing is a necessary step to ensure that faithful topological features (e.g. the correct number of components or holes) are present before computing Euler Characteristics. 
Random rotations of the data ahead of pre-processing lead to negligible differences in ECT and thus suggest this pre-processing does not introduce artificial geometric features.

\section{Results}
We analyse the shape of organoid boundaries of experimental (2D) and synthetic (3D) data by first building a simplicial complex representation. 
We then compute the SECT of each organoid at each time point to obtain DETECT. When computing DETECT, we use $a=6$ for the 2D organoids, $a=15$ for the 3D organoids and $P$ to be 100 evenly spaces points in $[-a, a]$. We then compute an $m$-dimensional feature embedding ($m=100$ for the experimental data and $m=30$ for the smaller synthetic data set; both data sets use a Gaussian kernel with $\lambda=|N|\times |P|$).

\subsection{Regressing SECT to classical shape statistics}

We first compute the SECT of static images of experimental organoids and demonstrate that the SECT includes information conveyed by classical shape statistics. The classic shape statistics, diameter, the mean and max centroid distances, the equivalent diameter, the major and minor axis lengths and the area of the convex hull, quantify geometric properties of a 2D shape.
These statistics are widely used and invariant under translation and $O(2)$ actions. As each of these statistics is calculated for a static shape, i.e. for an organoid boundary at a fixed time frame, we compare these statistics to the SECT at fixed time frames.

\begin{figure}
    \centering
    \includegraphics[width=.8\textwidth]{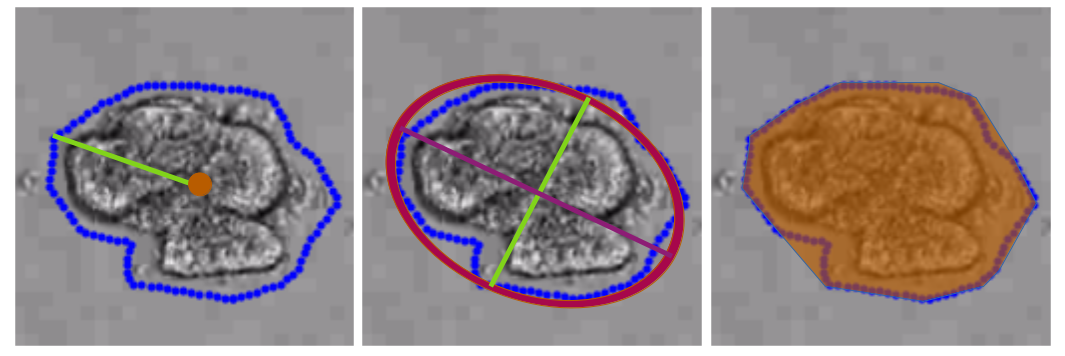}
    \caption{Left: The red point gives the centroid (point which minimises mean squared distance to boundary) of the organoid and the green line gives the distance to a boundary point. The maximal and mean lengths of such green lines give the max. and mean centroid distance. Centre: The red ellipse gives the ellipse with the best fit to the boundary. The purple line gives the major axis and the green line the minor axis. Right: the area of the convex hull (convex area) is visualised in opaque red.}
    \label{fig:regression_notions}
\end{figure}

To compare the aforementioned shape statistics with the SECT, we apply a standard linear regression model. The SECT of each organoid at each time is represented by a Nystroem feature embedding. We project the feature embedding to 50 dimensions using PCA \cite{pearson1901liii}. The PCA vectors give the independent variables, while the classic shape statistics listed above are viewed as the dependent variables. We pass the square-root values of the convex hull area to the regression model, as it has a squared relationship with the remaining metrics in the (default) symmetric cases. An illustration of the main notions is given in Figure \ref{fig:regression_notions}. We perform a 50-fold cross-validation for each metric and report mean coefficients of determination and standard deviations of the coefficient of determination in Table \ref{tab:reg_results}.

\begin{table}[]
    \centering
    \begin{tabular}{|c||c|c|}
    \hline
     & $R^2$-scores & std \\\hline\hline
    equivalent diameter & 0.880 & 0.073 \\\hline
    max. centroid distance & 0.916 & 0.041 \\\hline
    mean centroid distance & 0.894 & 0.067 \\\hline
    major axis length & 0.880 & 0.052 \\\hline
    minor axis length & 0.630 & 0.246 \\\hline
    perimeter & 0.972 & 0.024 \\\hline
    $\sqrt{\text{convex area}}$ & 0.868 & 0.105 \\\hline
    \end{tabular}
    \caption{Mean coefficients of determinations ($R^2$-scores) and their standard deviations (std) in a 50-fold cross-validation of a Linear regression in which the Nystroem features of the SECT give the independent variables and the 8 variables above give the dependent variables.}
    \label{tab:reg_results}
\end{table}

We find that the SECT regresses multiple classical shape statistics with high accuracy. We note that the diameter of an organoid should be proportional to both the major axis length and the maximum centroid distance. 

The SECT has high predictive accuracy of equivalent diameters and perimeter and, as a result, can also detect symmetry breaking. The lower accuracy of the minor axis length and convex area suggests that the SECT is more limited in its ability to capture the (mean) size of indentations, compared to detecting size and elongation. We remark that this limitation may be related to segmentation accuracy, and therefore, we consider organoid shapes with known segmentation (i.e. synthetic data) in Section \ref{sec:syn_results}.

\begin{figure}
    \centering
    \includegraphics[width=.7\textwidth]{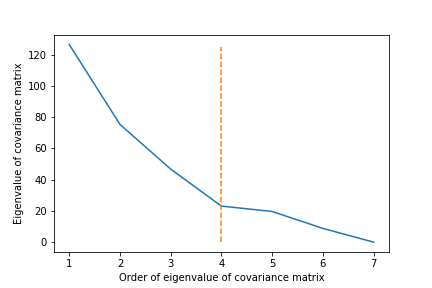}
    \caption{The eigenvalues ($y$-axis) of the covariance matrix of the classical shape descriptors in decreasing order (indices of eigenvalues on $x$-axis). The data was standardised in each component before the covariance matrix was computed. The first four eigenvalues (to the left of the dashed line) account for 90.5\% of the variance in the data.}
    \label{fig:cca_svd}
\end{figure}

In addition to the above regression analysis, we can decompose the covariance matrix of the aforementioned classical shape statistics by its singular values. We remark that we standardise the data in each feature before computing the covariances. We observe that the first four principal components of the classical shape statistics explain over 90\% of the variance in this data set (see Figure \ref{fig:cca_svd}). We then perform a canonical correlation analysis (CCA) \cite{wegelin2000survey} between these four principal components and the PCA-transformed SECT data. We find that the first four pairs of canonical variables have a perfect correlation score of 1.0 and conclude that the SECT can explain over 90\% of the variance in the classical shape statistics on the given data set.

\subsection{Classification of Organoids}

We focus on classifying p53-knock-out organoids into treated and untreated groups. Before training a classifier, we seek to exclude some organoids where the segmentation does not accurately trace the organoid boundary in the video. In these cases, the segmented boundary is significantly larger than the true organoid boundary. Therefore, we exclude organoids whose radius grows by a factor of more than two. For p53-knock-out organoids, this procedure excludes two out of 98 organoids. Of the remaining 96 organoids, 55 are untreated.

We use random forest classification \cite{breiman2001random} to classify p53-knock-out organoids into untreated and VC-treated experimental groups. Random forest classification then trains an ensemble of decision trees trained on random subsets of the Nystroem-transformed DETECT data. The trees classify data points by majority vote. We use the \texttt{scikit-learn} \cite{scikit-learn} implementation of random forest classification and optimise the hyper-parameters of the maximum tree depth, the minimum number of samples allowed to define a split and the minimum number of samples per tree leaf by cross-validation grid search (\texttt{GridSearchCV} in \texttt{scikit-learn}). Based on this optimisation, maximum tree depth is five, minimum number of samples to define a split is five and minimum number of samples per tree leaf is three.
The 5-fold cross-validation for these parameters gives a mean classification accuracy of 68.8\% with a standard deviation of 2.7\%. We remark that setting the number of Nystroem features to $m=500$ increases the mean accuracy further to 70.0\% but also increases the standard deviation to 7.5\%. The higher standard deviation suggests that setting $m=500$ could result in overfitting.

The accuracy of classification results based on DETECT exceeds those based on all classical statistics (e.g. area and perimeter) which give a mean classification accuracy of 60.5\% and a standard deviation of 4.8\% when we use the pipeline and cross-validation method described above. Classification based on DETECT also exceeds the baseline accuracy associated with guessing, which is 57.4\% as there are slightly more untreated than treated organoids in our data set.
Further, we have shown that combining DETECT with machine learning can distinguish organoids treated with valproic acid and GSK3 inhibitor based on quantification of their shape dynamics as well as regress out classical shape statistics.

\subsection{Distinguishing shape evolution 3D}\label{sec:syn_results}

Finally, we apply our methodology to the synthetic data generated by the model of Yan, Kostorum and Lowengrub \cite{yan2018three}. As described in Section \ref{sec:data}, this data set contains 9 organoids and thus is too small to  apply linear regression or random forest classification. We therefore only report the first two principal components of the Nystroem-transformed DETECT signatures. These outputs demonstrate that our methods generalise well to 3D shapes and identify important structure in the synthetic data, for which we know the ground truth.

\begin{figure}
    \centering
    \begin{subfigure}{0.47\textwidth}
         \centering
         \includegraphics[width=\textwidth]{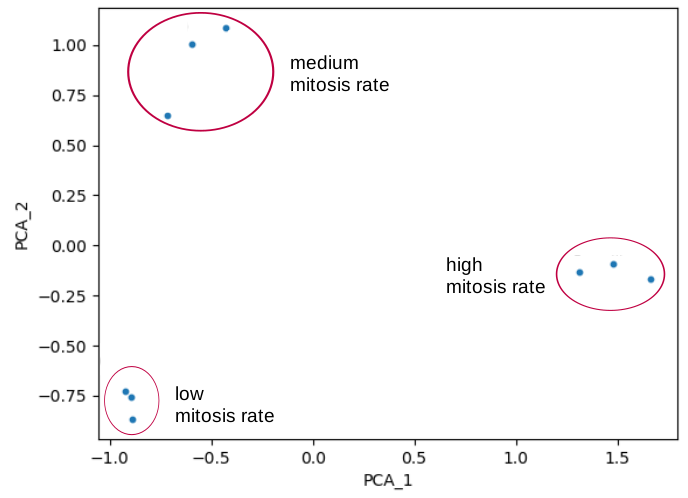}
         \caption{}
    \end{subfigure}%
    \begin{subfigure}{0.47\textwidth}
         \centering
         \includegraphics[width=0.99\textwidth]{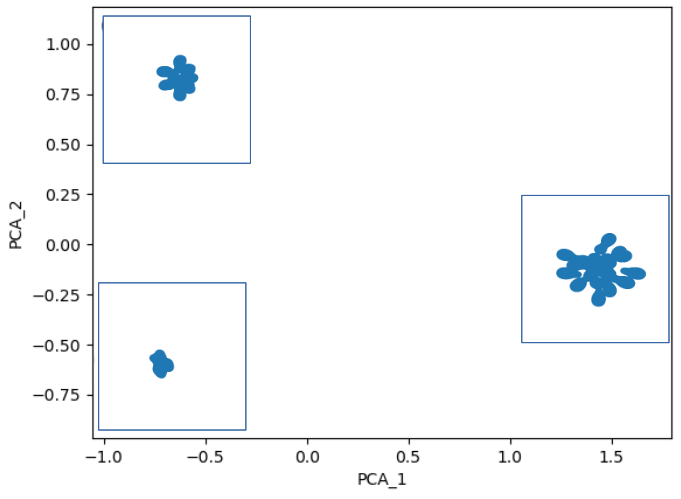}
         \caption{}
    \end{subfigure}
    \caption{(A) The first two principal components of the Nystroem-transformed DETECT for the 3D analysis. We see that these Nystroem features of the DETECT signatures cluster by mitosis rate, which is the dominant signal in the data. (B) Example of 2D projections of an organoid with given mitosis rate at the final time-point.}
    \label{fig:kpca}
\end{figure}

This analysis, visualised in Figure \ref{fig:kpca}, shows that organoids cluster together by their mitosis rates. This behaviour is consistent with watching the videos for these 9 organoids. We see that low-mitosis-rate organoids exhibit little growth and virtually no symmetry breaking (see Figure \ref{fig:syn_time_course}). Medium-mitosis-rate organoids show a little more growth than low-proliferation organoids and notable buckling of their boundary (see Figure \ref{fig:syn_time_course}). Finally, high-mitosis-rate organoids exhibit a large degree of growth and strong development of protrusions. In fact, the development of protrusions is so pronounced that several protrusions disconnect from the main organoid at later time points (see Figure \ref{fig:syn_time_course}).

The principal components in Figure \ref{fig:kpca} therefore appear to pick up the major signal in the synthetic 3D data set. We hypothesise that the first principal component is proportional to the size of the organoid. Similarly, we conjecture that the second principal component corresponds to the geometric complexity of the organoids. In particular, low-mitosis-rate organoids have the lowest geometric complexity while medium-mitosis-rate exhibit significant symmetry breaking. The high-mitosis-rate organoids lie in between the two former groups of organoids in terms of geometric complexity, as their protrusion development is so pronounced that protrusions disconnect. The resulting connected tissues are relatively spherical.

\section{Discussion}

In this work, we have introduced a new technique from the field of topological data analysis, for detecting temporal shape changes with the Euler Characteristic Transform (DETECT). We have highlighted its utility by studying organoid morphology. We first showed that several classical shape descriptors, including the diameter, the mean and maximum centroid distances, the equivalent diameter, the major and minor axis lengths and the area of the convex hull, can be regressed from the Smooth Euler Characteristic Transform (SECT) with high accuracy.
We applied DETECT, with kernel approximation methods and random forests, to a data set of experimental p53 knock-out mouse small intestine organoids, and showed that our approach can distinguish VC-treated  organoids from untreated organoids.  
We remark that this integration with kernel approximations enables larger data sets to be analysed because when kernel methods are used the runtime complexity of ECT can be reduced from being cubic to approximately linear in the number of data points. 
We highlighted this methodology generalises to 3D by applying DETECT to an \emph{in silico} data set derived from a continuum model of organoid growth \cite{yan2018three}. DETECT enables the synthetically generated organoids to be clustered according to mitosis rate (one of the parameters which were varied in the synthetic data).

In the future, we aim to extend our findings - both in 2D and 3D - to data sets of different types of organoids (derived from different organs, with different genetic backgrounds and/or cultured under different conditions). 
We aim also to study information loss between 3D data and their 2D projections. We will accomplish this by first considering synthetic data generated from 3D mechanistic models \cite{yan2018three}, which neglect certain biophysical processes (e.g. the effects of gravity, the production of extracellular matrix, mechanical stress) \cite{yan2018three}. This will enable us to focus on establishing relationships between the DETECT signatures and the values of key model parameters (e.g., cell proliferation and differentiation rates).

We plan to extend this analysis to other types of morphological data that do not have regularised and smooth boundaries. For example, there are no random perturbations in the synthetic data studied here. In practice, data sets analysed by the ECT and its extensions may be noisier than the data sets analysed in this paper.
The ECT (and, by extension, its variants, including DETECT) are not stable with respect to small perturbations. Theoretical work studying the stability properties of the ECT and extensions would provide a significant step forward for their application to noisy morphological data sets.

\section*{Acknowledgements}
We thank Huaming Yan, Anna Konstorum and John Lowengrub for helpful discussions and for providing us with data generated by their model of organoid growth.
HAH gratefully acknowledges funding from a Royal Society University Research Fellowship.
LM, HMB and HAH are members of the Centre for Topological Data Analysis, which is funded by the EPSRC grant `New Approaches to Data Science: Application Driven Topological Data Analysis' \href{https://gow.epsrc.ukri.org/NGBOViewGrant.aspx?GrantRef=EP/R018472/1}{\texttt{EP/R018472/1}}. FYZ was funded by the Ludwig Institute for Cancer Research and an EPSRC Life Sciences Interface Doctoral Training Centre Fellowship, EP/F500394/1. XQ was funded by Ludwig Cancer Institute, China Scholarship Council, and Oxford University.
For the purpose of Open Access, the authors have applied a CC BY public copyright licence to any  Author Accepted Manuscript (AAM) version arising from this submission.

\printbibliography

@article{scikit-learn,
 title={Scikit-learn: Machine Learning in {P}ython},
 author={Pedregosa, F. and Varoquaux, G. and Gramfort, A. and Michel, V.
         and Thirion, B. and Grisel, O. and Blondel, M. and Prettenhofer, P.
         and Weiss, R. and Dubourg, V. and Vanderplas, J. and Passos, A. and
         Cournapeau, D. and Brucher, M. and Perrot, M. and Duchesnay, E.},
 journal={Journal of Machine Learning Research},
 volume={12},
 pages={2825--2830},
 year={2011}
}

@article{tang2022topological,
  title={A topological data analytic approach for discovering biophysical signatures in protein dynamics},
  author={Tang, Wai Shing and da Silva, Gabriel Monteiro and Kirveslahti, Henry and Skeens, Erin and Feng, Bibo and Sudijono, Timothy and Yang, Kevin K and Mukherjee, Sayan and Rubenstein, Brenda and Crawford, Lorin},
  journal={PLoS computational biology},
  volume={18},
  number={5},
  pages={e1010045},
  year={2022},
  publisher={Public Library of Science San Francisco, CA USA}
}

@article{pearson1901liii,
  title={LIII. On lines and planes of closest fit to systems of points in space},
  author={Pearson, Karl},
  journal={The London, Edinburgh, and Dublin philosophical magazine and journal of science},
  volume={2},
  number={11},
  pages={559--572},
  year={1901},
  publisher={Taylor \& Francis}
}

@article{yan2018three,
  title={Three-dimensional spatiotemporal modeling of colon cancer organoids reveals that multimodal control of stem cell self-renewal is a critical determinant of size and shape in early stages of tumor growth},
  author={Yan, Huaming and Konstorum, Anna and Lowengrub, John S},
  journal={Bulletin of mathematical biology},
  volume={80},
  number={5},
  pages={1404--1433},
  year={2018},
  publisher={Springer}
}

@article{kumar2012sampling,
  title={Sampling methods for the Nystr{\"o}m method},
  author={Kumar, Sanjiv and Mohri, Mehryar and Talwalkar, Ameet},
  journal={The Journal of Machine Learning Research},
  volume={13},
  number={1},
  pages={981--1006},
  year={2012},
  publisher={JMLR. org}
}

@article{Amezquita2021,
author = {Erik J Amezquita and Michelle Y Quigley and Tim Ophelders and Jacob B. Landis and Daniel Koenig and Elizabeth Munch and Daniel H Chitwood},
doi = {10.1093/insilicoplants/diab033},
journal = {in silico Plants},
publisher = {Oxford University Press ({OUP})},
title = {Measuring hidden phenotype: Quantifying the shape of barley seeds using the Euler Characteristic Transform},
year = {2021}
}

@inproceedings{dey2018protein,
  title={Protein classification with improved topological data analysis},
  author={Dey, Tamal K and Mandal, Sayan},
  booktitle={18th International Workshop on Algorithms in Bioinformatics (WABI 2018)},
  year={2018},
  organization={Schloss Dagstuhl-Leibniz-Zentrum fuer Informatik}
}

@inproceedings{jiang2020weighted,
  title={The Weighted Euler Curve Transform for Shape and Image Analysis},
  author={Jiang, Qitong and Kurtek, Sebastian and Needham, Tom},
  booktitle={Proceedings of the IEEE/CVF Conference on Computer Vision and Pattern Recognition Workshops},
  pages={844--845},
  year={2020}
}

@inproceedings{martino2018supervised,
  title={Supervised approaches for protein function prediction by topological data analysis},
  author={Martino, Alessio and Rizzi, Antonello and Mascioli, Fabio Massimo Frattale},
  booktitle={2018 International joint conference on neural networks (IJCNN)},
  pages={1--8},
  year={2018},
  organization={IEEE}
}

@article{gameiro2015topological,
  title={A topological measurement of protein compressibility},
  author={Gameiro, Marcio and Hiraoka, Yasuaki and Izumi, Shunsuke and Kramar, Miroslav and Mischaikow, Konstantin and Nanda, Vidit},
  journal={Japan Journal of Industrial and Applied Mathematics},
  volume={32},
  number={1},
  pages={1--17},
  year={2015},
  publisher={Springer}
}

@article{de2007coverage,
  title={Coverage in sensor networks via persistent homology},
  author={De Silva, Vin and Ghrist, Robert},
  journal={Algebraic \& Geometric Topology},
  volume={7},
  number={1},
  pages={339--358},
  year={2007},
  publisher={Mathematical Sciences Publishers}
}

@article{beers2022stability,
  title={Stability of topological descriptors for neuronal morphology},
  author={Beers, David and Harrington, Heather A and Goriely, Alain},
  journal={arXiv preprint arXiv:2211.09058},
  year={2022}
}

@article{marsh2022algebra,
  title={Algebra, geometry and topology of ERK kinetics},
  author={Marsh, Lewis and Dufresne, Emilie and Byrne, Helen M and Harrington, Heather A},
  journal={Bulletin of Mathematical Biology},
  volume={84},
  number={12},
  pages={1--50},
  year={2022},
  publisher={Springer}
}

@article{williams2000using,
  title={Using the Nystr{\"o}m method to speed up kernel machines},
  author={Williams, Christopher and Seeger, Matthias},
  journal={Advances in neural information processing systems},
  volume={13},
  year={2000}
}

@article{ishihara2022topological,
  title={Topological morphogenesis of neuroepithelial organoids},
  author={Ishihara, Keisuke and Mukherjee, Arghyadip and Gromberg, Elena and Brugu{\'e}s, Jan and Tanaka, Elly M and J{\"u}licher, Frank},
  journal={Nature Physics},
  pages={1--7},
  year={2022},
  publisher={Nature Publishing Group}
}

@article{wang2021statistical,
  title={A statistical pipeline for identifying physical features that differentiate classes of 3D shapes},
  author={Wang, Bruce and Sudijono, Timothy and Kirveslahti, Henry and Gao, Tingran and Boyer, Douglas M and Mukherjee, Sayan and Crawford, Lorin},
  journal={The Annals of Applied Statistics},
  volume={15},
  number={2},
  pages={638--661},
  year={2021},
  publisher={Institute of Mathematical Statistics}
}

@article{gidea2017topology,
  title={Topology data analysis of critical transitions in financial networks},
  author={Gidea, Marian},
  journal={arXiv preprint arXiv:1701.06081},
  year={2017}
}

@article{leibon2008topological,
  title={Topological structures in the equities market network},
  author={Leibon, Gregory and Pauls, Scott and Rockmore, Daniel and Savell, Robert},
  journal={Proceedings of the National Academy of Sciences},
  volume={105},
  number={52},
  pages={20589--20594},
  year={2008},
  publisher={National Acad Sciences}
}

@article{kramar2013persistence,
  title={Persistence of force networks in compressed granular media},
  author={Kram{\'a}r, Miroslav and Goullet, Arnaud and Kondic, Lou and Mischaikow, Konstantin},
  journal={Physical Review E},
  volume={87},
  number={4},
  pages={042207},
  year={2013},
  publisher={APS}
}

@article{kramar2014quantifying,
  title={Quantifying force networks in particulate systems},
  author={Kram{\'a}r, Miroslav and Goullet, Arnaud and Kondic, Lou and Mischaikow, Konstantin},
  journal={Physica D: Nonlinear Phenomena},
  volume={283},
  pages={37--55},
  year={2014},
  publisher={Elsevier}
}

@article{perea2015sw1pers,
  title={SW1PerS: Sliding windows and 1-persistence scoring; discovering periodicity in gene expression time series data},
  author={Perea, Jose A and Deckard, Anastasia and Haase, Steve B and Harer, John},
  journal={BMC bioinformatics},
  volume={16},
  number={1},
  pages={1--12},
  year={2015},
  publisher={BioMed Central}
}

@article{delgado2014skeletonization,
  title={Skeletonization and partitioning of digital images using discrete morse theory},
  author={Delgado-Friedrichs, Olaf and Robins, Vanessa and Sheppard, Adrian},
  journal={IEEE transactions on pattern analysis and machine intelligence},
  volume={37},
  number={3},
  pages={654--666},
  year={2014},
  publisher={IEEE}
}

@inproceedings{boche2013signal,
  title={Signal analysis with frame theory and persistent homology},
  author={Boche, Holger and Guillemard, Mijail and Kutyniok, Gitta and Philipp, Friedrich},
  booktitle={The Conference of Sampling Theory and Applications, SampTA’13},
  volume={1},
  pages={1},
  year={2013}
}

@inproceedings{chung2009persistence,
  title={Persistence diagrams of cortical surface data},
  author={Chung, Moo K and Bubenik, Peter and Kim, Peter T},
  booktitle={International Conference on Information Processing in Medical Imaging},
  pages={386--397},
  year={2009},
  organization={Springer}
}

@article{breiman2001random,
  title={Random forests},
  author={Breiman, Leo},
  journal={Machine learning},
  volume={45},
  number={1},
  pages={5--32},
  year={2001},
  publisher={Springer}
}

@misc{preproc,
  title = {Delaunay+Voronoi on a sphere},
  howpublished = {\url{ https://www.redblobgames.com/x/1842-delaunay-voronoi-sphere/
}},
  note = {Accessed: 2022-08-02}
}

@article{kanari2018topological,
  title={A topological representation of branching neuronal morphologies},
  author={Kanari, Lida and D{\l}otko, Pawe{\l} and Scolamiero, Martina and Levi, Ran and Shillcock, Julian and Hess, Kathryn and Markram, Henry},
  journal={Neuroinformatics},
  volume={16},
  number={1},
  pages={3--13},
  year={2018},
  publisher={Springer}
}

@article{kim2020comparison,
  title={Comparison of cell and organoid-level analysis of patient-derived 3D organoids to evaluate tumor cell growth dynamics and drug response},
  author={Kim, Seungil and Choung, Sarah and Sun, Ren X and Ung, Nolan and Hashemi, Natasha and Fong, Emma J and Lau, Roy and Spiller, Erin and Gasho, Jordan and Foo, Jasmine and others},
  journal={SLAS DISCOVERY: Advancing the Science of Drug Discovery},
  volume={25},
  number={7},
  pages={744--754},
  year={2020},
  publisher={Sage Publications Sage CA: Los Angeles, CA}
}

@Article{Byrne2020,
  author  = {Byrne, H},
  journal = {Interface Focus},
  title   = {Three Dimensional Biological Cultures and Organoids},
  year    = {2020},
  number  = {20200014},
  volume  = {10},
}

@article{kretzschmar2016organoids,
  title={Organoids: modeling development and the stem cell niche in a dish},
  author={Kretzschmar, Kai and Clevers, Hans},
  journal={Developmental cell},
  volume={38},
  number={6},
  pages={590--600},
  year={2016},
  publisher={Elsevier}
}

@article{Turner2014,
  title={Persistent homology transform for modeling shapes and surfaces},
  author={Turner, Katharine and Mukherjee, Sayan and Boyer, Doug M},
  journal={Information and Inference: A Journal of the IMA},
  volume={3},
  number={4},
  pages={310--344},
  year={2014},
  publisher={Oxford University Press}
}

@Article{Curry2019,
  author  = {Curry, Justin AND Mukherjee, Sayan AND Turner, Katharine},
  journal = {arXiv: 1805.09782v2},
  title   = {How Many Directions Determine A Shape and Other Sufficiency Results for Two Topological Transforms},
  year    = {2019},
}

@Article{Crawford2020,
  author  = {Crawford, Lorin AND Monod, Anthea AND Chen, Andrew X. AND Mukherjee, Sayan AND Rabadan, Raul},
  journal = {Journal of the American Statistical Association},
  title   = {Predicting Clinical Outcomes in Glioblastoma: An Application of Topological and Functional Data Analysis},
  year    = {2020},
  number  = {531},
  pages   = {1139-1150},
  volume  = {115},
}

@article{ghrist2018persistent,
  title={Persistent homology and Euler integral transforms},
  author={Ghrist, Robert and Levanger, Rachel and Mai, Huy},
  journal={Journal of Applied and Computational Topology},
  volume={2},
  number={1},
  pages={55--60},
  year={2018},
  publisher={Springer}
}

@article{kassis2019orgaquant,
  title={OrgaQuant: human intestinal organoid localization and quantification using deep convolutional neural networks},
  author={Kassis, Timothy and Hernandez-Gordillo, Victor and Langer, Ronit and Griffith, Linda G},
  journal={Scientific reports},
  volume={9},
  number={1},
  pages={1--7},
  year={2019},
  publisher={Nature Publishing Group}
}

@article{gritti2021morgana,
  title={MOrgAna: accessible quantitative analysis of organoids with machine learning},
  author={Gritti, Nicola and Lim, Jia Le and Anla{\c{s}}, Kerim and Pandya, Mallica and Aalderink, Germaine and Martinez-Ara, Guillermo and Trivedi, Vikas},
  journal={Development},
  volume={148},
  number={18},
  pages={dev199611},
  year={2021},
  publisher={The Company of Biologists Ltd}
}

@article{wegelin2000survey,
  title={A survey of Partial Least Squares (PLS) methods, with emphasis on the two-block case},
  author={Wegelin, Jacob A},
  year={2000},
  publisher={Citeseer}
}

@article{hartung2014mathematical,
  title={Mathematical modeling of tumor growth and metastatic spreading: validation in tumor-bearing mice},
  author={Hartung, Niklas and Mollard, S{\'e}verine and Barbolosi, Dominique and Benabdallah, Assia and Chapuisat, Guillemette and Henry, Gerard and Giacometti, Sarah and Iliadis, Athanassios and Ciccolini, Joseph and Faivre, Christian and others},
  journal={Cancer research},
  volume={74},
  number={22},
  pages={6397--6407},
  year={2014},
  publisher={AACR}
}

@article{buske2012biomechanics,
  title={On the biomechanics of stem cell niche formation in the gut--modelling growing organoids},
  author={Buske, Peter and Przybilla, Jens and Loeffler, Markus and Sachs, Norman and Sato, Toshiro and Clevers, Hans and Galle, Joerg},
  journal={The FEBS journal},
  volume={279},
  number={18},
  pages={3475--3487},
  year={2012},
  publisher={Wiley Online Library}
}

@Article{Thalheim2018,
  author   = {Thalheim, Torsten and Quaas, Marianne and Herberg, Maria and Braumann, Ulf-Dietrich and Kerner, Christiane and Loeffler, Markus and Aust, Gabriela and Galle, Joerg},
  journal  = {Developmental Biology},
  title    = {Linking stem cell function and growth pattern of intestinal organoids},
  year     = {2018},
  issn     = {0012-1606},
  number   = {2},
  pages    = {254--261},
  volume   = {433},
  url      = {http://www.sciencedirect.com/science/article/pii/S0012160617303767},
}

\end{document}